\documentclass[aps,twocolumn,showpacs,preprintnumbers,prb]{revtex4}
\usepackage{graphicx}

\newcommand{\R}{\mathbf{r}}

\newcommand{\UP}{n_{\uparrow}}
\newcommand{\DN}{n_{\downarrow}}

\newcommand{\be}{\begin{equation}}
\newcommand{\ee}{\end{equation}}
\newcommand{\bea}{\begin{eqnarray}}
\newcommand{\eea}{\end{eqnarray}}
\newcommand{\bean}{\begin{eqnarray*}}
\newcommand{\eean}{\end{eqnarray*}}

\newenvironment{acknowledgment}{{\flushleft \bf Acknowledgments:}}{}

\begin{document}

\title{Kinetic energy density functionals from the Airy gas, with an application to\\ 
the atomization kinetic energies of molecules} 
\author{Lucian A. Constantin and Adrienn Ruzsinszky}
\affiliation{Department of Physics and Quantum Theory Group,
Tulane University, New Orleans, LA 70118}

\date{\today}


\begin{abstract}
We construct and study several semilocal density functional approximations for the positive 
Kohn-Sham kinetic energy density. These functionals fit the kinetic energy density of the Airy gas 
and they can be accurate for integrated kinetic energies of atoms, molecules, jellium clusters and 
jellium surfaces. We find that these functionals are the most accurate ones for atomization 
kinetic 
energies of molecules and for fragmentation of jellium clusters. We also report that local and 
semilocal kinetic energy functionals can show "binding" 
when the density of a spin unrestricted Kohn-Sham calculation is used.         
\end{abstract}

\maketitle

\section{Introduction}
\label{sec1}
\noindent

The positive Kohn-Sham (KS) \cite{KS} kinetic energy (KE) density of noninteracting electrons
\begin{equation} 
\tau(\R)=\frac{1}{2}\sum^N_i |\nabla\phi_i(\R)|^2,
\label{e1b}
\end{equation}
is an exact functional of the occupied orbitals $\{\phi_i\}$.
Density functional approximations to the noninteracting kinetic energy $T_s[\UP,\DN]=\int d\R 
\tau(\R)$ can simplify and 
speed up by orders of magnitude 
any KS self-consistent calculation \cite{note1}. 
(Here $\UP(\R)$ and $\DN(\R)$ are the spin densities.)
However, in spite of important and hard work done 
in this 
direction \cite{KJTH}, no actual approximation has reached chemical accuracy. 
 
The simplest model of an edge electron gas is the Airy gas, where any electron feels
a linear 
effective potential \cite{KM}, and thus the normalized one-particle eigenfunctions are 
proportional to 
the Airy function. The effective finite-linear-potential model gives remarkably good 
results for 
the jellium surface problem \cite{SMF,SS3}. 
However, the KE density 
derived in this 
approximation \cite{Ba} does not recover the correct second-order gradient expansion 
KE density \cite{Ki,BJC} 
and has an unphysical oscillating behavior in the limit of slow density variations \cite{DG},
being a poor approximation
for atoms \cite{GB}.

The positive KE 
density of the Airy gas was studied by Vitos \emph{et. al} \cite{Vi1}, and they 
derived 
a generalized gradient approximation (GGA) density functional for 
$\tau(\R)$.
(This approximation is denoted in this paper by VJKS GGA.)
They showed 
that the poor behavior of the kinetic energy density derived in the
linearized-potential approximation \cite{Ba} is mainly due to a Laplacian 
term 
that arises naturally in the Airy gas model. Thus, the Laplacian term, even if it 
integrates to zero and does not affect the integrated KE, is an important tool in 
developing density functionals not only for 
the KE but also for the exchange-correlation (xc) energy \cite{Vi1,PC}.
The VJKS GGA KE density functional 
fits the Airy gas KE  density and is a good model for the KE density 
of the jellium surfaces,
but for atoms and molecules it diverges to $-\infty$ at the nuclei, due to the behavior of 
the Laplacian term. The integrated kinetic energies are at a Thomas-Fermi 
\cite{TF} level of accuracy, reducing considerably the error of the
linearized-potential approximation \cite{Ba}. 

A jellium surface is the simplest model of a metallic surface. Self-consistent 
local-spin-density (LSD) calculations \cite{LK} for this model provided early evidence 
that density functionals may work. But wavefunction-based methods, like Fermi 
hypernetted 
chain \cite{KK} and Diffusion Monte Carlo (DMC) of Ref. \cite{AC} predicted 
low-density surface xc energies 
about 40\% larger than those from LSD.
Recent refined DMC estimates \cite{WHFGG}, and calculations in the random phase approximation 
\cite{PE,PCP} and beyond it \cite{PP2,CPDGLP,CPT}, agree with the popular xc semilocal 
density 
functionals, showing that the jellium surface can not only be accurately described in 
the context 
of density functional theory, but can also be an important model used 
to develop new density functionals.

The exchange energy density of the Airy gas \cite{KM,Vi2,AM05} and the xc jellium surface energies 
\cite{AM05,PRCVSCZB} were employed 
in the construction of 
accurate xc GGA's for solids. (See Refs. \cite{Vi2,AM05,PRCVSCZB}). A simple xc GGA functional 
depends only on spin 
densities and their gradients and can not 
describe accurately both solids and atoms \cite{PCSB}. 
However, a Laplacian-level 
xc meta-GGA \cite{PC}, that depends nontrivially on spin densities and their gradients and 
Laplacians, 
can be accurate for atoms, molecules, solids and surfaces.

In this paper, we derive 
several GGA KE functionals from the Airy gas and jellium surfaces and we find them accurate 
for 
atomization KE energies of molecules and for fragmentation of jellium clusters. 
Our functionals, constructed similarly to that of Ref. \cite{Vi1}, recover the 
second-order 
gradient expansion of the
integrated KE, have the right behavior of the KE density in the tail of the density, and 
fit the kinetic energy density of the Airy gas.

The paper is organized as follows. In section \ref{sec2},
we construct our KE 
functionals. In section 
\ref{sec3} 
we test the functionals for atoms, jellium clusters, jellium surfaces and molecules. In section 
\ref{sec4}, we summarize our
conclusions.

\section{Laplacian-dependent GGA kinetic energy functionals}
\label{sec2}
\noindent

The positive kinetic energy density of the local Airy gas (LAG) is \cite{Vi1}
\begin{equation}
\tau^{LAG}(z)=-\frac{3}{5}n(z)v_{eff}(z)+\frac{1}{5} \nabla^2 n(z),
\label{e1}
\end{equation}
where $v_{eff}(z)$ is the effective potential and $n(z)$ is the density of the Airy gas. 
(Unless otherwise stated, atomic units are used throughout, i.e.,
$e^2=\hbar=m_e=1$.)
Alternatively, Eq. (\ref{e1})
can be written \cite{Vi1} using the Thomas-Fermi kinetic energy density 
$\tau^{TF}=(3/10)(3\pi^2)^{2/3}n^{5/3}$:
\begin{equation}
\tau^{LAG}(z)=\tau^{TF}(z)P(z)+\frac{1}{5}\nabla^2 n(z),
\label{e2}
\end{equation}
where 
\begin{equation}
P(z)=-\frac{2Bz}{(3\pi^2)^{2/3}n(z)^{2/3}},
\label{e3}
\end{equation}
and $B$ is the slope of the linear effective potential. $P(z)$ is a smooth function of the reduced 
density gradient
\begin{equation}
s(\R)=|\nabla n(\R)|/[2k_F(\R)n(\R)], 
\label{eeb3}
\end{equation}
where $k_F(\R)=(3\pi^2 n(\R))^{1/3}$ is the Fermi wavevector. (The dimensionless density 
gradient $s(\R)$ measures the variation of the density over a Fermi wavelength 
$\lambda_F=2\pi/k_F$.)
Thus, Vitos \emph{et. al} \cite{Vi1} proposed the following GGA KE 
density functional
\begin{equation}
\tau^{VJKS}(\R)=\tau^{TF}(\R)P^{VJKS}(s(\R))+\frac{1}{5}\nabla^2 n(\R),
\label{e4}
\end{equation}
where 
\begin{equation}
P^{VJKS}(s)=\frac{1+0.8944s^2-0.0431s^6}{1+0.6511s^2+0.0431s^4}
\label{e5}
\end{equation}
fits $P(z)$ for the Airy gas model.
Eq. (\ref{e4}) recovers the exact 
KE density of the 
von Weizs\"{a}cker functional \cite{vW} 
$ |\nabla n|^2 /(8n) = (5/3) \tau^{TF} s^2 $
for an exponentially decaying density (see Ref. \cite{note5}), but 
for a slowly-varying density behaves as 
$\tau^{TF}(1+0.2433s^2+O(s^4))+\frac{1}{5}\nabla^2 n(z)$ and violates
the second-order gradient expansion (GE2) of the KE density \cite{Ki,BJC} 
\begin{equation}
\tau^{GE2}=\tau^{TF}(1+\frac{5}{27}s^2)+\frac{1}{6}\nabla^2 n.
\label{e6}
\end{equation}

Let us consider the following arbitrary partition of Eq. (\ref{e2}) for the Airy gas model
\begin{equation}
\tau^{LAG}(z)=\tau^{TF}(z)F(z,\beta)+\beta\nabla^2 n.
\label{e7}
\end{equation}
Eqs. (\ref{e2}) and (\ref{e7}) give
\begin{equation}
F(z,\beta)=P(z)+\frac{[(1/5)-\beta]\nabla^2 n(z)}{\tau^{TF}(z)}.
\label{e8}
\end{equation}
$F(z,\beta)$ is a smooth function of
the reduced gradient $s$ for any $\beta >1/8$, and it can be accurately approximated by the 
following expression
\begin{equation}
F^{CR}(s,\beta)=\frac{1+(a_1+5/27)s^2+a_2s^4+a_3s^6-a_4s^8}{1+a_1s^2+a_5s^4+\frac{3}{40\beta 
-5}a_4s^6},
\label{e9}
\end{equation}
where $a_1,a_2,a_3,a_4$ and $ a_5$ are parameters that depend on $\beta$. Eq. (\ref{e9}) recovers 
the 
terms $1+(5/27)s^2$ for a slowly-varying density, but the second-order gradient expansion
of the KE density additionally requires that $\beta = 1/6$.
In the tail, where the density decays exponentially, Eqs. (\ref{e7}) and 
(\ref{e9}) give the correct KE density of 
the von Weizs\"{a}cker functional.

When $\beta=1/5$, $F(z,\beta=1/5)=P(z)$ and we define a GGA (A$\frac{1}{5}$) similar with the 
one in Ref. \cite{Vi1}
\begin{equation}
\tau^{A\frac{1}{5}}(\R)=\tau^{TF}(\R)F^{CR}(s(\R),\beta=1/5)+\frac{1}{5}\nabla^2 n(\R),
\label{e10}
\end{equation}
where the fitting parameters are shown in Table \ref{ta1b}. 

\begin{table}[htbp]
\footnotesize
\caption{ Parameters of the enhancement factor $F^{CR}(s,\beta)$ for various GGAs.} 
\begin{tabular}{|l|l|l|l|}
   \multicolumn{1}{c}{ } &
   \multicolumn{1}{c}{ } &
   \multicolumn{1}{c}{ }\\  \hline
   & A$\frac{1}{5}$-GGA &  A$\frac{1}{6}$-GGA & A0.185-GGA  \\  \hline
$a_1$ & 1.122609 & 1.301786 & 1.293576 \\  \hline
$a_2$ & 0.900085 & 3.715282 & 2.161116 \\  \hline
$a_3$ & -0.227373 & 0.343244 & -0.144896 \\  \hline
$a_4$ & 0.014177 & 0.032663 & 0.025505 \\  \hline
$a_5$ & 0.731298 & 2.393929 & 1.444659 \\  \hline
\end{tabular}
\label{ta1b}
\end{table}

When $\beta=1/6$, we define a GGA (A$\frac{1}{6}$) that recovers the second-order gradient 
expansion KE density
\begin{equation}
\tau^{A\frac{1}{6}}(\R)=\tau^{TF}(\R)F^{CR}(s(\R),\beta=1/6)+\frac{1}{6}\nabla^2 n(\R),
\label{e11}
\end{equation}
where the fitting parameters are shown in Table \ref{ta1b}. 

\begin{figure}
\includegraphics[width=\columnwidth]{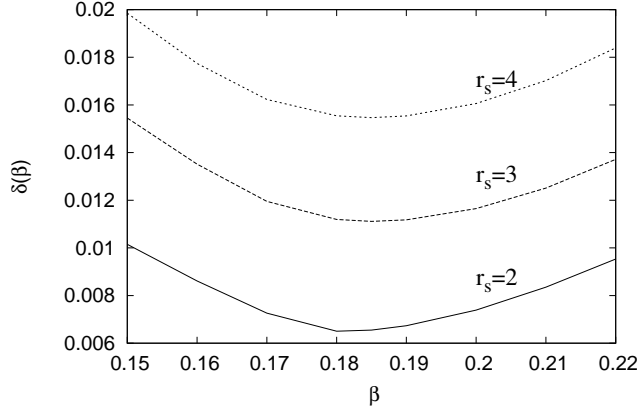}
\caption{ The quality factor $\delta(\beta)$ versus $\beta$, for $\tau^{approx}$ given by Eqs.
(\ref{e7}) and (\ref{e9}), for
the jellium surfaces with bulk parameters $r_s=2,3,$ and $4$. We use LSD KS orbitals and 
densities \cite{LK,MP}.}
\label{f1}
\end{figure}

The Airy gas is the simplest edge electron gas and does not include curvature corrections 
that are present at the edge surfaces (see 
Fig. 2 of Ref. \cite{KM}). Thus in order to find an optimum value of $\beta$ for jellium 
surfaces, 
let us define the quality factor (similarly to Refs. \cite{Vi1}
and \cite{GAA})
\begin{equation}
\delta(\beta)=\int d\R\; |\tau^{approx}(\R,\beta)-\tau(\R)|/\int d\R\;\tau(\R),
\label{e12}
\end{equation}
where $\tau^{approx}$ is an approximation of the positive Kohn-Sham KE density $\tau$. [See Eq. 
(\ref{e1b})]. 
We apply the quality factor to jellium surfaces using numerical LSD Kohn-Sham orbitals and 
densities \cite{LK,MP}. The integration 
was done from
$z_{\min}=-2.75\lambda_F$ to $z_{\max}=2\lambda_F$, where $\lambda_F=2\pi/k_F$ is the bulk Fermi 
wavelength
, for
several values of bulk parameter $r_s$. (Here $r_s=(9\pi/4)^{1/3}/k_F$ is the radius of a 
sphere which contains on
average one electron, and $k_F$ is the bulk Fermi wavevector.)
For $\tau^{approx}$ we use Eqs. (\ref{e7}) and (\ref{e9}). Thus, for values of $\beta$ between 
0.15 and 0.22,
we accurately fit $F(z,\beta)$ with the Pad\'{e} approximation of Eq.
(\ref{e9}), and we calculate $\delta(\beta)$. Fig. \ref{f1} shows that $\delta(\beta)$ is minimum 
for 
$\beta\approx 0.185$ for semi-infinite jellium surfaces with $r_s=2,3,$ and $4$.
%
%

So from our jellium surface analysis we define the following GGA (A0.185)
that also fits the kinetic energy density of the Airy gas
\begin{equation}
\tau^{A0.185}(\R)=\tau^{TF}(\R)F^{CR}(s(\R),\beta=0.185)+0.185\nabla^2 n(\R),
\label{e14}
\end{equation}
where the fitting parameters are shown in Table \ref{ta1b}. 

\begin{figure}
\includegraphics[width=\columnwidth]{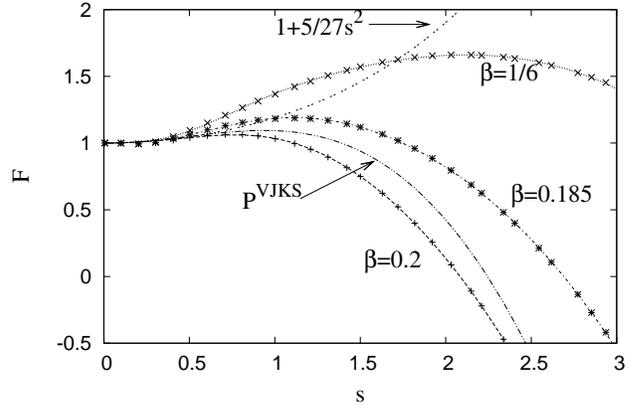}
\caption{ The exact function $F(z,\beta)$ shown with points $(F(z,\beta),s(z))$ for some 
discrete $z$, and parametrized function 
$F^{CR}(s,\beta)$ shown with lines
for $\beta=1/5, 1/6$ and 0.185, versus the reduced gradient $s$, for the Airy gas model.
Also shown are the enhancement factor $(1+5/27s^2)$ and $P^{VJKS}(s)$.}
\label{f2}
\end{figure}

In Fig. \ref{f2} we show the exact function $F(z,\beta)$ and the fitting function 
$F^{CR}(s,\beta)$ 
versus the scaled density gradient $s$, for 
$\beta=1/5, 1/6$ and 0.185 respectively. Up to $s=3$, the exact functions $F$ and the 
parametrized 
ones can not be distinguished. (We note that $s$ values bigger than 3 are
found in the tail of an atom or molecule, where the electron density is negligible.)
$P^{VJKS}(s)$ overestimates $P(z)=F(z,\beta=1/5)$ until $s\approx 3$ and underestimates 
$P(z)$ for $3\leq 
s\leq 10$. 
%
%

\begin{figure}
\includegraphics[width=\columnwidth]{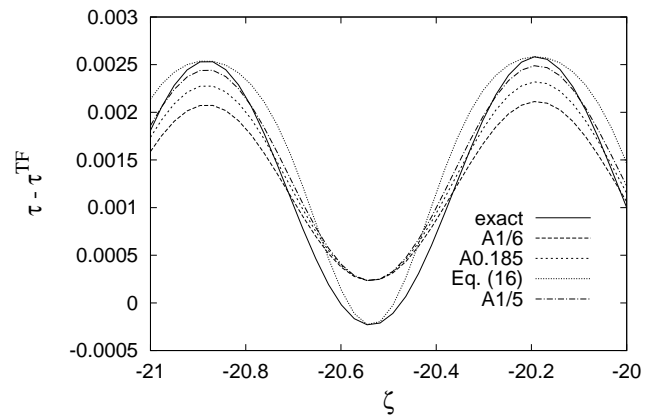}
\caption{ $\tau-\tau^{TF}$ versus $\zeta$ for the Airy gas model. The edge is at $\zeta=0$. 
The integrations of $\tau-\tau^{TF}$ over the complete Friedel oscillation shown in figure are: 
$T_s^{\rm{exact}}-T_s^{TF}=8.30\times 10^{-4}$, $T_s^{A\frac{1}{5}}-T_s^{TF}=9.58\times 
10^{-4}$, 
$T_s^{A\frac{1}{6}}-T_s^{TF}=8.26\times 
10^{-4}$,
$T_s^{A0.185}-T_s^{TF}=8.98\times 10^{-4}$, and $T_s^{lin}-T_s^{TF}=9.89\times 10^{-4}$. 
VJKS GGA, not shown in the figure, gives an integrated value of $10.09\times 10^{-4}$.}
\label{f3}
\end{figure}

Far from the edge of the Airy gas, the density has Friedel oscillations \cite{KM}. These 
oscillations are well described by the kinetic energy density of the linear potential approximation 
\cite{Ba} that in the slowly-varying density regime reduces to \cite{DG}
\begin{equation}
\tau^{lin}=\tau^{TF}+\frac{5}{72}\frac{(\nabla n)^2}{n}+\frac{1}{12}\frac{(\nabla 
n)^2}{n}\sin(\frac{2(3\pi^2)^{1/3}n^{4/3}}{|\nabla n|}).
\label{e15}
\end{equation}
The third term represents quantum oscillations and has an unphysical behavior when $\nabla 
n\rightarrow 0$. In Fig. \ref{f3} we show $\tau-\tau^{TF}$ versus $\zeta$,
for a slowly-varying Airy gas density. The edge is at $\zeta=0$. 
[$\zeta=(2B)^{1/3}z$ is the scaled spatial coordinate
for the Airy gas.]
The Friedel oscillations are well 
described by Eq. (\ref{e15}). But even if $\tau^{A\frac{1}{6}}-\tau^{TF}$ is the worst kinetic 
energy 
density 
shown in the figure, its integration over a period of the Friedel oscillations is almost exact. 
Thus 
$\tau^{A\frac{1}{6}}$, that behaves as $\tau^{GE2}$ in this limit, is the best approximation 
for the 
integrated 
KE, whereas $\tau^{lin}$ gives the worst integrated KE.   
%
%

\section{Tests of our GGA kinetic energy functionals}
\label{sec3}

In this section we test 
our functionals for various systems. In the calculations we use the spin-scaling relation 
\cite{OP}
\begin{equation}
\tau_\sigma([n_{\sigma}],\R)=(1/2)\tau([n=2n_{\sigma}],\R),
\label{e16}
\end{equation}
where $n_{\sigma}$ is the density of the electrons with spin $\sigma$. ($\sigma=\uparrow$ or 
$\downarrow$.)

\subsection{Integrated kinetic energies of atoms, jellium clusters and jellium surfaces}
\label{ss1}
In Table \ref{ta1} we show the accuracy of $T^{TF}_s$, $T^{VJKS}_s$, 
$T^{GE2}_s$, $T^{GE4}_s$,
$T^{A\frac{1}{5}}_s$, $T^{A\frac{1}{6}}_s$, and $T^{A0.185}_s$ for
atoms, jellium clusters and jellium surfaces (similarly as Table I of Ref. \cite{PC}). The error
displayed in this table is
\begin{eqnarray}
& \mathrm{Error}=\frac{1}{2}\mathrm{``m.a.r.e. atoms"}+
\frac{1}{4}\mathrm{``m.a.r.e. clusters"}\nonumber\\
& +\frac{1}{4}\mathrm{``m.a.r.e. LDM(N=8)"},
\label{e17}
\end{eqnarray}
where ``m.a.r.e. atoms" is the mean absolute relative error
(m.a.r.e.) of the integrated kinetic energy of
50 atoms and ions (listed in Ref. \cite{PC}), ``m.a.r.e. clusters"
is the m.a.r.e. of $2e^{-}$, $8e^{-}$, $18e^{-}$, $20e^{-}$,
$34e^{-}$, $40e^{-}$, $58e^{-}$, $92e^{-}$, and $106e^{-}$ neutral
spherical jellium
clusters (with bulk parameter $r_{s}=3.93$ which corresponds to
Na), and ``m.a.r.e. LDM(N=8)" is the m.a.r.e. of the KE of
N=8 jellium spheres for $r_s$ = 2, 4, and 6, calculated in the
liquid drop model \cite{PC} (LDM)
\begin{equation}
T^{LDM}_{s}=(3/10)k^{2}_{F}N+\sigma_{s}N^{2/3}4\pi r^{2}_{s},
\label{e18}
\end{equation}
where 
$k_F$ is the bulk Fermi wavevector, and $\sigma_{s}$ is the surface KE.
The exact LDM value is computed with the exact $\sigma_s$
(using LSD orbitals). 
Because the relative errors of surface kinetic energies
are much larger than those of the atoms and spherical jellium clusters,
we use the LDM approach for calculating the jellium surface
KE errors (as in Ref. \cite{PC}); LDM gives m.a.r.e. comparable to that of
atoms and clusters (see Table \ref{ta1}).
We use analytic Hartree-Fock densities and orbitals \cite{CR11} for atoms and ions,  
and numerical Kohn-Sham densities and orbitals for jellium clusters (using the optimized potential 
method (OPM) \cite{KP})  and jellium semi-infinite surfaces (using LSD xc potential).
\begin{table}[htbp]
\footnotesize
\caption{ Mean absolute relative error (m.a.r.e.) of
kinetic energies of 50 atoms and ions (see Ref. \cite{PC}), of neutral spherical
jellium Na clusters ($2e^{-}$, $8e^{-}$, $18e^{-}$,
$20e^{-}$, $34e^{-}$, $40e^{-}$, $58e^{-}$, $92e^{-}$, and $106e^{-}$)
and of jellium surfaces (with $r_{s}=2$, $r_{s}=4$, and $r_{s}=6$)
incorporated into the liquid drop model (LDM) for a
jellium sphere with N=8 electrons
(see Eq. (\ref{e18})). Also shown is the total error given by Eq. (\ref{e17}).}
\begin{tabular}{|l|l|l|l|l|}
   \multicolumn{1}{c}{ } &
   \multicolumn{1}{c}{ } &
   \multicolumn{1}{c}{ } &
   \multicolumn{1}{c}{ } &
   \multicolumn{1}{c}{ }\\  \hline
 & m.a.r.e.& m.a.r.e.& m.a.r.e.&  \\  
 & atoms & clusters & LDM(N=8) &  Error (Eq. (\ref{e17})) \\  \hline
$T^{TF}_{s}$ & 0.0842 & 0.0439 & 0.0810 & 0.0733 \\  \hline
$T^{VJKS}_{s}$ & 0.0399 & 0.0465 & 0.0754 & 0.0504 \\  \hline
$T^{GE2}_{s}$ & 0.0112 & 0.0099 & 0.0330 & 0.016 \\  \hline
$T^{GE4}_{s}$ & 0.0251 & 0.0176 & 0.0170 & 0.0212 \\  \hline
$T^{A\frac{1}{5}}_s$ & 0.0626 & 0.0566 & 0.0879 & 0.067 \\  \hline 
$T^{A\frac{1}{6}}_s$ & 0.0789 & 0.0154 & 0.0177 & 0.048 \\  \hline 
$T^{A0.185}_s$ & 0.0083 & 0.0249 & 0.0535 & 0.024 \\  \hline
\end{tabular}
\label{ta1}
\end{table}

$\tau^{VJKS}$, $\tau^{A\frac{1}{5}}$, $\tau^{A\frac{1}{6}}$ and $\tau^{A0.185}$ are constructed 
to model the KE density of the Airy gas, 
but only $\tau^{A\frac{1}{6}}$ recovers the second-order gradient expansion of the KE density. 
The 
difference between 
$\tau^{A\frac{1}{5}}$ and $\tau^{VJKS}$ is given mainly by the quality of fitting the function 
$P(z)$ of Eq. 
(\ref{e3}). (See Fig. \ref{f2}.) $\tau^{A0.185}$ includes effects of density
variations near jellium surfaces because of our optimization of the Laplacian coefficient. 
In Table \ref{ta1} we see that $T^{A\frac{1}{6}}_s$ is very accurate (comparable with the 
fourth-order 
gradient expansion) for jellium systems and gives an overall
error smaller than $T^{A\frac{1}{5}}_s$ and $T^{VJKS}_s$.
$T^{A0.185}_s$ is accurate for atoms and gives an overall error comparable with the 
fourth-order gradient 
expansion one ( see also Table 1 of Ref. \cite{PC}). 
\subsection{Integrated atomization kinetic energy for a set of molecules}
\label{ss2}
In Table \ref{ta2} we present the atomization kinetic energies for the molecules used in Refs. 
\cite{PC,IEMS}. We observe that $T_s^{A\frac{1}{5}}$ keeps the right sign for all the 
molecules and has practically the same 
mean absolute error as the Thomas-Fermi functional. In Ref. \cite{IEMS} it was shown that 
the 
Thomas-Fermi KE functional gives better atomization kinetic energies than all the 
other tested semilocal functionals. $T_s^{A0.185}$ is accurate for atoms and molecules, and 
gives 
the smallest mean absolute error for the atomization energies presented in Table \ref{ta2}. We 
also 
show that the PBE-like semilocal functional of Ref. \cite{TW}, whose parameters are fitted to 
atoms, works worse than the Thomas-Fermi functional and all the semilocal functionals derived 
from the Airy gas.  
\begin{table}[htbp]
\footnotesize
\caption{ Integrated atomization kinetic energy
( KE atoms - KE molecule, in a. u.) for 
the set of molecules used in Refs. \cite{PC,IEMS}.
The kinetic energies were calculated
using the PROAIMV code with Kohn-Sham orbitals given by the Gaussian 2000 code
(with the uncontracted $6-311+G(3df,2p)$ basis set, Becke 1988 exchange functional
\cite{Becke},
and Perdew-Wang correlation functional \cite{PW91}).
The last line shows the mean absolute errors (m.a.e.). Here $T^{TW}_s$ is the the GGA of Ref. 
\cite{TW} 
with the parameters $k=0.8438$ and $\mu=0.2319$.}
\begin{tabular}{|l|l|l|l|l|l|l|l|l|}
   \multicolumn{1}{c}{ }&
   \multicolumn{1}{c}{ }&
   \multicolumn{1}{c}{ }&
   \multicolumn{1}{c}{ }&
   \multicolumn{1}{c}{ }&
   \multicolumn{1}{c}{ }&
   \multicolumn{1}{c}{ } \\  \hline
& $T^{\mathrm{exact}}_{s}$ & $T^{TF}_{s}$ & $T^{VJKS}_{s}$ &
$T^{GE2}_s$ & $T^{A\frac{1}{5}}_s$ & $T^{A\frac{1}{6}}_s$ & $T^{A0.185}_s$ & $T^{TW}_s$ \\  
\hline
$\mathrm{H}_2$ & -0.150 & -0.097 & -0.086 & -0.114 & -0.080 & -0.114 & -0.096 & -0.108 \\  \hline
$\mathrm{HF}$  & -0.185 & -0.305 & -0.369 & -0.186 & -0.422 & -0.173 & -0.311 & -0.226 \\  \hline
$\mathrm{H}_{2}\mathrm{O}$ & -0.304 & -0.308 & -0.455 & -0.136 & -0.531 & -0.169 & -0.369 & -0.209 \\  
\hline
$\mathrm{CH}_4$  & -0.601 & -0.737 & -0.907 & -0.571 & -0.972 & -0.618 & -0.813 & -0.649 \\  \hline
$\mathrm{NH}_3$  & -0.397 & -0.231 & -0.457 & -0.060 & -0.525 & -0.165 & -0.364 & -0.155 \\  \hline
$\mathrm{CO}$  & -0.298 & -0.323 & -0.580 & -0.085 & -0.678 & -0.181 & -0.456 & -0.203 \\  \hline
$\mathrm{F}_2$  & -0.053 & 0.128 & 0.013 & 0.282 & -0.050 & 0.269 & 0.093 & 0.223 \\  \hline
$\mathrm{HCN}$  & -0.340 & -0.1835 & -0.539 & 0.079 & -0.644 & -0.097 & -0.399 & -0.071 \\  \hline
$\mathrm{N}_2$  & -0.158 & 0.344 & -0.046 & 0.565 & -0.134 & 0.321 & 0.069 & 0.412 \\  \hline
$\mathrm{CN}$  & -0.431 & -0.215 & -0.539 & 0.005 & -0.631 & -0.168 & -0.424 & -0.129 \\  \hline
$\mathrm{NO}$  & -0.268 & 0.092 & -0.215 & 0.330 & -0.313 & 0.176 & -0.094 & 0.198 \\  \hline
$\mathrm{O}_2$  & -0.100 & 0.106 & -0.089 & 0.335 & -0.177 & 0.286 & 0.030 & 0.239 \\  \hline\hline
m.a.e. &               & 0.177 & 0.133 & 0.311 & 0.172 & 0.224 & 0.116 & 0.232 \\  \hline
\end{tabular}
\label{ta2}
\end{table}

\subsection{ Binding energy of the $\rm{N}_2$ molecule}
\label{ss3}
\begin{figure}
\includegraphics[width=\columnwidth]{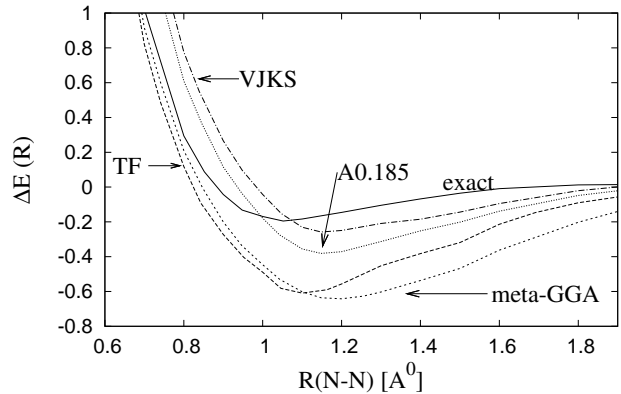}
\caption{ Binding energy ( $\Delta$E = E molecule - E atoms, in a.u.)  as a function of N-N 
distance
for the $\rm{N}_2$
molecule
using a nonrestricted Hartree-Fock calculation (with uncontracted 6-311+G(3df,2p) basis set).
The curve "meta-GGA" is the binding energy given by the Laplacian-level KE meta-GGA of Ref. 
\cite{PC}.
The Hartree-Fock density was used as input for orbital free KE functionals. $1\AA=1.8897 
a.u.$.}
\label{f4}
\end{figure}

In Fig. \ref{f4} we show the
binding energy of the $\rm{N}_2$ molecule as a function of the distance between the nuclei.
We use a spin unrestricted Hartree-Fock calculation in which 
the spin symmetry
breaks close to the Hartree-Fock equilibrium bond length. This helps the functionals to show an 
equilibrium length close to the exact. Figure \ref{f4} is in accord with the values for 
the $N_2$ 
molecule listed in Table \ref{ta2};  
all the semilocal functionals presented in the figure give 
bigger atomization kinetic energies than the exact calculation, thus showing a minimum in the 
total 
energy calculated with the  Hartree-Fock density. 

The unrestricted solution becomes energetically lower beyond
the Coulson-Fisher point \cite{FNGB} than the energy of the restricted solution, and 
spin symmetry
breaking for the $\rm{N}_2$ molecule can be achieved by mixing the highest occupied and
lowest unoccupied orbitals \cite{LH}.
For a spin-restricted calculation, the orbital-free KE functionals listed in Table II do not show 
an equilibrium 
point, thus the 
spin-breaking symmetry \cite{GL,PSB,PGH} and the spin-scaling relations \cite{OP} play 
an important role in describing 
stretched molecules, and they need to be taken into account in the orbital-free codes.     
%
%

\subsection{ Tests of the kinetic energy density}
\label{ss4}
\begin{figure}
\includegraphics[width=\columnwidth]{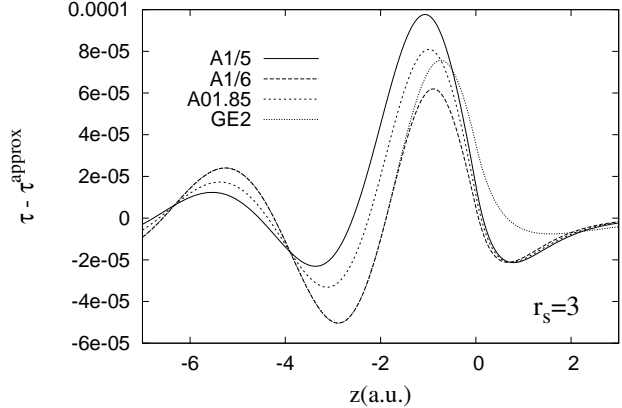}
\caption{$\tau(z)-\tau^{approx}(z)$, where $\tau^{approx}$ is $\tau^{A\frac{1}{5}}$,
$\tau^{A\frac{1}{6}}$, $\tau^{A0.185}$, and $\tau^{GE2}$ respectively, versus $z$, for
a jellium surface of bulk parameter $r_s=3$. 
The surface is at $z=0$, the jellium is at $z\leq0$ and the vacuum is at $z>0$.
The surface kinetic energies are:
$\sigma^{exact}_s=-703\; \mathrm{erg}/\mathrm{cm}^{2}$, 
$\sigma^{A\frac{1}{5}}_s=-869\; \mathrm{erg}/\mathrm{cm}^{2}$, 
$\sigma^{A\frac{1}{6}}_s=-690\; \mathrm{erg}/\mathrm{cm}^{2}$, 
$\sigma^{A0.185}_s=-788\; \mathrm{erg}/\mathrm{cm}^{2}$, and 
$\sigma^{GE2}_s=-762\; \mathrm{erg}/\mathrm{cm}^{2}$.
VJKS GGA, not plotted in the figure, gives
$\sigma^{VJKS}_s=-837\; \mathrm{erg}/\mathrm{cm}^{2}$.
($1\mathrm{hartree}/\mathrm{bohr}^2=1.557\times 10^6 \mathrm{erg}/\mathrm{cm}^{2}$.) 
We use LSD KS orbitals and
densities \cite{LK,MP}.
}
\label{f5}
\end{figure}

In Fig. \ref{f5} we show the kinetic energy density of our functionals at a jellium surface. 
Though $\tau^{A0.185}$ has the smallest overall error, $\tau^{A\frac{1}{6}}$ gives the most 
accurate surface 
kinetic energy because it is accurate near the surface and it can 
almost exactly damp the Friedel oscillations far from the surface (see Fig. \ref{f3}).
%

\begin{figure}
\includegraphics[width=\columnwidth]{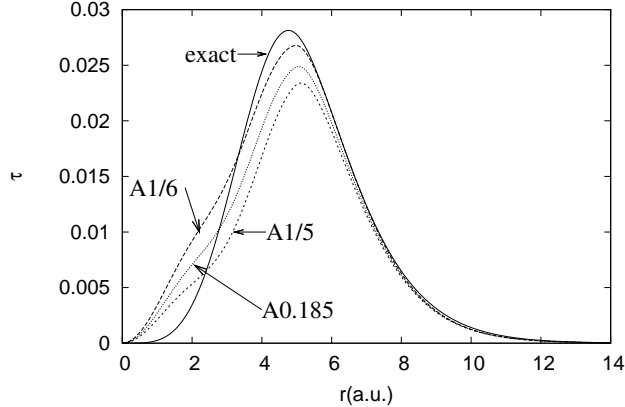}
\caption{ Kinetic energy density versus radial distance $r$, for the $2e^-$ jellium
cluster (with bulk parameter
$r_s=3.93$). The area under the curve is the kinetic energy: $T^{exact}_s=0.114$ a.u.,
$T^{A\frac{1}{5}}_s=0.098$ a.u., $T^{A\frac{1}{6}}_s=0.121$ a.u. and $T^{A0.185}_s=0.108$ a.u.. 
VJKS GGA, not 
shown in 
the
figure, gives $T^{VJKS}_s=0.101$ a.u..}
\label{f6}
\end{figure}

In Fig. \ref{f6} we show the kinetic energy densities of our functionals for the $2e^-$ Na jellium
cluster. Here the exact curve is the von  Weizs\"{a}cker \cite{vW} KE density. We see that all 
three functionals ($\tau^{A\frac{1}{5}}$, $\tau^{A\frac{1}{6}}$, and $\tau^{A0.185}$) recover 
the 
exact curve 
in the tail 
of 
the density, as expected.
%

\subsection{ Large-$Z$ asymptotic behavior}
\label{ss5}
The non-interacting kinetic energy of the neutral atoms has the following asymptotic expansion 
\cite{En88,LBCP}:
\begin{equation}
T_s=c_0Z^{7/3}+c_1Z^2+c_2Z^{5/3},
\label{e19}
\end{equation} 
where $Z$ is the atomic number, and $c_0=0.768745$, $c_1=-1/2$, and $c_2=0.2699$. In Ref. 
\cite{LBCP} the authors propose an accurate method to extract these coefficients for any KE 
functional. In Table \ref{ta3} we present the large-$Z$ asymptotic behavior of our functionals.
All the functionals listed in Table \ref{ta3} are exact for systems with uniform density, such 
that we expect that they have the exact Thomas-Fermi coefficient $c_0=0.768745$. (Similarly 
with Ref. 
\cite{LBCP}, we do not have enough data points to extract $c_0$ accurately.) 
$T^{A0.185}_s$ and $T^{VJKS}_s$, the functionals that give the most accurate atomization 
kinetic 
energies,
have reasonable large-$Z$ asymptotic behaviors.
\begin{table}[htbp]
\footnotesize
\caption{ The coefficients of the asymptotic expansion of Eq. (\ref{e19}) for several semilocal 
functionals. The fitting method is the same as in Ref. \cite{LBCP}. We use 
OPM \cite{KP} densities.  }
\begin{tabular}{|l|l|l|l|}
   \multicolumn{1}{c}{ } &
   \multicolumn{1}{c}{ } &
   \multicolumn{1}{c}{ }\\  \hline
   & $c_0$ & $c_1$ & $c_2$  \\  \hline
Exact & 0.768745 & -0.500000 & 0.269900   \\  \hline
$T^{GE2}_s$ & 0.768745 & -0.536197 & 0.335992 \\  \hline
$T^{TW}_s$ & 0.768745 & -0.507979 & 0.291815 \\  \hline
$T^{A\frac{1}{5}}_s$ & 0.768745 & -0.532065 & 0.229370 \\  \hline
$T^{A\frac{1}{6}}_s$ & 0.768745 & -0.439745 & 0.392152 \\  \hline
$T^{A0.185}_s$ & 0.768745 & -0.491080 & 0.302999 \\  \hline
$T^{VJKS}_s$ & 0.768745 & -0.507589 & 0.225358 \\  \hline
\end{tabular}
\label{ta3}
\end{table}

\subsection{ Fragmentation of jellium clusters}
\label{ss6}
Let us consider the disintegration of the $106e^-$ 
neutral spherical jellium Na cluster into smaller closed-shell
jellium spheres:
\begin{eqnarray}
(106e^-)\longrightarrow 
n_1(92e^-)+n_2(58e^-)+n_3(40e^-)+n_4(34e^-)\nonumber\\
+n_5(20e^-)+n_6(18e^-)+n_7(8e^-)+n_8(2e^-),
\label{e20}
\end{eqnarray}
where $n_1,...,n_8$ are positive integers, and 
$92n_1+58n_2+40n_3+34n_4+20n_5+18n_6+8n_7+2n_8=106$.
We define the disintegration KE as  
\begin{equation}
\rm{DKE}=\rm{KE\;of\;initial\;cluster\; -\; KE\; of\; the\; fragments}.
\label{e21}
\end{equation}
%
\begin{figure}
\includegraphics[width=\columnwidth]{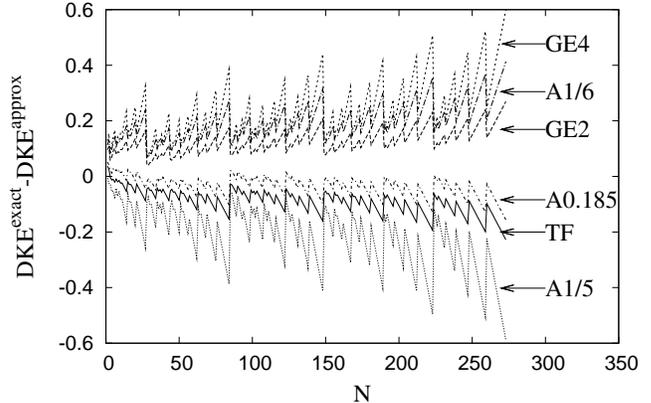}
\caption{ Error of disintegration KE ($\rm{DKE}^{\rm{exact}}-\rm{DKE}^{\rm{approx}}$) for 273 
configurations described by Eq. (\ref{e20}). The first point $N=1$, corresponds to 
$(106e^-)\rightarrow (92e^-)+(8e^-)+3\times (2e^-)$, and the last point N=273 corresponds to 
$(106e^-)\rightarrow 53\times (2e^-)$. We use OPM-KS orbitals and densities.
Mean absolute errors are: $\rm{m.a.e}^{\rm{GE4}}=0.247$,
$\rm{m.a.e}^{\rm{A\frac{1}{5}}}=0.218$,
$\rm{m.a.e}^{\rm{GE2}}=0.128$,
$\rm{m.a.e}^{\rm{TF}}=0.092$,
$\rm{m.a.e}^{\rm{A0.185}}=0.039$, and
$\rm{m.a.e}^{\rm{A\frac{1}{6}}}=0.196$. VJKS GGA, not shown in the figure, has 
$\rm{m.a.e}^{\rm{VJKS}}=0.165$.
} 
\label{f7}
\end{figure}
%
In Fig. \ref{f7} we show $\rm{DKE}^{\rm{exact}}-\rm{DKE}^{\rm{approx}}$ for 273 processes
described by Eq. (\ref{e20}), for several KE functionals. We see that our functional 
$T_s^{A0.185}$
is very accurate, improving over $T_s^{TF}$ for all the configurations. $T_s^{A\frac{1}{6}}$ is 
close to, 
but
better than the fourth-order gradient expansion $T_s^{GE4}$. Overall, this figure agrees
well with
the atomization KE of molecules reported in Table \ref{ta2}, showing an important link between
jellium spheres and molecules. These results and the liquid drop model (see Eq. (17) of
Ref.\cite{LBCP}) suggest that the TF functional gives a good balance between jellium surface KE 
and
jellium curvature KE. This balance, that is important in atomization and disintegration processes, 
is improved by the A0.185-GGA functional.

\section{Conclusions}
\label{sec4}
\noindent

In this paper we have studied several semilocal KE density functionals derived from the Airy gas. 
These functionals, that depend trivially on the Laplacian of the density, do not satisfy several 
important 
constraints. Their kinetic energy densities are not always positive, 
and they implicitly violate the important constraint $\tau^{approx}\geq\tau^{W}$ (here
$\tau^{W}$ is the von Weizs\"{a}cker KE density) and  diverge to $-\infty$ at the
nucleus of an atom.

However, such functionals can be accurate for the integrated KE of jellium surfaces and jellium 
clusters (e.g. 
$T^{A\frac{1}{6}}_s$), and of atoms and molecules (e.g. $T^{A0.185}_s$), when we use realistic 
densities ( from 
KS 
calculations). More importantly, they 
are the most accurate KE density functionals, to our knowledge, for the integrated atomization 
kinetic energies of molecules and for the fragmentation of jellium clusters.
These functionals may also be useful for quasi-realistic densities (e.g. a superposition of 
free-atom Kohn-Sham densities), but they are not accurate enough for orbital-free calculations. 

We have also presented a spin-unrestricted Hartree-Fock calculation for the stretched $N_2$ 
molecule that explains the $N_2$ atomization kinetic energies displayed in Table \ref{ta2}, and 
that shows equilibrium lengths for many semilocal functionals. Thus, this work suggests that the 
spin-symmetry breaking and the spin scaling relations can be important tools 
in orbital-free approaches.


\begin{acknowledgment}
We thank Professor John P. Perdew for many valuable discussions and suggestions.
L.A.C. acknowledges NSF support (Grant No. DMR05-01588).
\end{acknowledgment}


\begin{thebibliography}{[Vo]}
\bibitem{note1}
An orbital-free kinetic energy density functional can be used in solving the spin-dependent Euler 
equations, or perhaps applied to a superposition of atomic
Kohn-Sham or Hartree-Fock densities.
\end{thebibliography}
\end{document}